\documentclass[12pt]{iopart}
\usepackage{epsfig}
\usepackage{iopams, setstack}

\newcommand{\bra}[1]{\left\langle#1\right|}
\newcommand{\hc}{\mathrm{h.c.}}
\newcommand{\idh}{\frac{\rmi}{\hbar}}
\newcommand{\ket}[1]{\left|#1\right\rangle}
\newcommand{\one}{\mathrm{1d}}
\newcommand{\openone}{1 \!\!\!\; \mathrm{l}}
\newcommand{\Hcond}{H_{\mathrm{cond}}}
\newcommand{\IL}{{{\mathcal I}_d}}
\newcommand{\ILsub}{{_\IL}}

\newcommand{\ground}{\ket{\downarrow}}
\newcommand{\groundbra}{\bra{\downarrow}}
\newcommand{\excited}{\ket{\uparrow}}
\newcommand{\excitedall}{\ket{\uparrow_1 \ldots \uparrow_D}}
\newcommand{\excitedbra}{\bra{\uparrow}}

\newcommand{\nodetect}{\ket{0}}
\newcommand{\nodetectproj}{\nodetect \! \bra{0}}

\begin{document}

\title[Quantum mechanical detector model for moving, spread-out
particles]%
{Investigation of a quantum mechanical detector model for
moving, spread-out particles}

\author{G C Hegerfeldt$^1$, J T Neumann$^1$ and L S Schulman$^2$} 
\address{$^1$ Institut f\"ur Theoretische Physik, Universit\"at
  G\"ottingen, Friedrich-Hund-Platz 1, 37077 G\"ottingen, Germany}
\address{$^2$ Physics Department, Clarkson University, Potsdam, New
  York 13699-5820, USA}
\eads{\mailto{hegerf@theorie.physik.uni-goettingen.de},
  \mailto{neumann@theorie.physik.uni-goettingen.de},
  \mailto{schulman@clarkson.edu}}

\begin{abstract}
  We investigate a fully quantum mechanical spin model for the
  detection of a moving particle. This model, developed in earlier
  work, is based on a collection of spins at fixed locations and in a
  metastable state, with the particle locally enhancing the coupling
  of the spins to an environment of bosons.  The appearance of bosons
  from particular spins signals the presence of the particle at the
  spin location, and the first boson indicates its arrival. The
  original model used discrete boson modes. Here we treat the
  continuum limit, under the assumption of the Markov property, and
  calculate the arrival-time distribution for a particle to reach a
  specific region.
\end{abstract}

\submitto{\JPA}


\section{Introduction}

Until recently, in time-of-flight measurements for particles or atoms
the quantum nature of the center-of-mass motion usually played
no role since the particles or atoms were very fast. However, the
advance of cooling techniques has made it possible to create ultracold
gases in a trap and produce very slow atoms, e.g., by opening the
trap. For these low velocities the quantum nature of the
center-of-mass motion of an atom can have noticeable quantum
effects, as the remarkable experiments of Szriftgiser et.\ al.\ have
shown \cite{sgad1996}. In the simplest quantum mechanical formulation
of a time-of-flight measurement one would create a particle at $t=0$
with a localized but extended wave function and then ask for the
arrival time of the particle at some distant point. Repeating this one
would get an arrival-time distribution that would depend on the
particle's wave function. Similarly one might ask for passage or
transit times through a region. Such questions, and more generally the
role of time in quantum mechanics, have attracted much interest in
recent years \cite{mme2002, lss1997}.

But how should one measure the arrival time of a particle or atom at
some particular point and what should the resulting distribution look
like? Allcock \cite{ga1969} made an \emph{ad hoc} model of an arrival-time
measurement using an imaginary step potential, which leads to an
`absorption' of the wave packet; he then identified the absorption
rate with the arrival-time distribution. In general, this
distribution will not be normalized since part of the wave packet will
be reflected from the imaginary step potential rather than being
absorbed. Also, part of the wave packet may penetrate the step to some
depth before being absorbed, thus causing a detection `delay.' As
Allcock noticed, decreasing one effect will typically enlarge the
other.

Kijowski proposed physically motivated axioms from which he derived an
`ideal' arrival-time distribution for a free quantum particle coming
from one direction \cite{jk1974}. The resulting distribution agrees
with an `approximate distribution' proposed heuristically by Allcock.
This distribution has been related \cite{rg1997} to the arrival-time
operator of Aharonov and Bohm \cite{ab_toa1961} (for more on the
latter see \cite{oru1999} and references therein). No measurement
procedure for the distribution was proposed, and its status,
properties and generalizations are still being critically discussed in
the literature; see e.g.\ \cite{crl2002, emnr2003, crl2005}.

Halliwell \cite{jjh1999} employed a detection model based on a single
spin coupled to a boson bath, a greatly simplified version of a
general quantum mechanical detector model that was proposed in
\cite{gs1990} and elaborated in \cite{lss1991, lss1997}. Working
in one space dimension and using Bloch equations he arrived at a
Schr\"odinger equation with an imaginary potential, thus giving a
basis for Allcock's approach.

An operational and realistic laser-based approach to the arrival-time
problem was investigated in \cite{dambo2002, hsm2003, bn2003b,
  dambo2003, hsmn2004, rdnmh2004, hhm2005}. This approach proposes to
measure the arrival-time by means of laser induced fluorescence
\cite{dambo2002}.  The idea is to consider a two-level atom with
center-of-mass motion, to illuminate some region of space with a laser,
and to take the detection time of the first fluorescence photon as the
arrival time of the atom at the (sharp) onset of the laser. In this
approach one has to deal with the typical problems of delay due to the
time needed for pumping and decay of the excited state. There is also
reflection without detection when the atom is reflected from the laser
beam in the ground state without emitting a photon. Yet, interesting
results could be derived. In the limit of a weak laser, there is
almost no reflection but a strong delay due to the weak pumping to the
upper energy level of the two-level system; dealing with this delay by
means of a deconvolution, one recovers the flux at the position of the
onset of the laser from the first-photon distribution
\cite{dambo2002}. On the other hand, in the limit of strong pumping,
reflection becomes dominant and the first-photon distribution is
clearly not normalized; normalizing it via the operator normalization
method of Brunetti and Fredenhagen, which preserves the bilinear
structure of the distribution \cite{bf2002}, one recovers Kijowski's
arrival-time distribution from the first-photon distribution
\cite{hsm2003}. In this way, Kijowski's axiomatic distribution can be
related to a particular measuring process. Further, in a certain limit
it is possible to derive a closed one-channel equation for the ground
state governing the first-photon distribution \cite{bn2003b}. This
equation contains an in general complex potential which becomes purely
imaginary for zero laser detuning. In this way the fluorescence model
makes a connection to Allcock's \emph{ad hoc} ansatz of an imaginary
potential.

In the fluorescence model there is a back-reaction of the measurement
on the center-of-mass motion of the atom, and this might cause
deviations from an ideal distribution. It thus seems a good idea to
use a measurement procedure that does not interact directly with the
particle through its internal degrees of freedom, but rather to regard
the particle only as a catalyst for a transition in a detector or its
associated environment.

Just such a detection model was developed in \cite{lss1997, gs1990,
  lss1991}. The model consists of a three-dimensional array of $D$
spins (the `detector') with ferromagnetic interaction. In the presence
of a homogeneous magnetic field, and for sufficiently low temperature,
all spins are aligned with the field. Reversing the magnetic field
suddenly, such that the spins cannot follow the reversal, one can
produce a metastable state of this compound spin system. The spins are
weakly coupled to a bath of bosons. There is a particle to be detected
and its effect on the collection of spins is to strongly enhance the
spin-boson coupling when the particle's wave function overlaps that of
a detector spin.  Thus when the particle is close to a spin this spin
flips much faster by virtue of the increased coupling to the bath. By
means of the ferromagnetic interaction, this in turn triggers the
subsequent spontaneous flipping of all spins even in the absence of
the particle.  In this way, the single spin flip is amplified to a
macroscopic event and the associated bosons can be measured. The
details of the amplification process and the probability of false
detection due to spontaneous spin flips were considered in
\cite{lss1997, gs1990, lss1991}. The motion of the particle, whose
presence induces the first spin flip, was treated classically in the
calculations. In the following, we will concentrate on the full
quantum description of this first spin flip and the quantum mechanical
aspects of the particle's motion, and comment only briefly on
processes internal to the detector.

In this paper we investigate this detector model in the limit of
continuous boson modes, under the condition that the spin-boson
interaction satisfies the Markov property (see (\ref{markovianproperty})), 
and use it to determine the arrival-time 
distribution of a spatially spread-out particle.
It turns out that one
is again led to a Schr\"odinger equation with an imaginary potential
and the corresponding arrival-time distribution is similar to that of
the fluorescence model. In this detector model there is also a
back-reaction on the particle of interest. In order to eliminate this
back-reaction we discuss the idea of decreasing the spin-bath coupling
while simultaneously increasing the number of spins. It is shown that
even in the limit when the spin-bath coupling goes to zero and the
number of spins to infinity, there remains a back-reaction.

The plan of the paper is as follows. In Section \ref{model} the
detector model is reviewed, and in Section \ref{direct_approach} the
arrival-time distribution obtained from a simplified version of the
model is calculated by means of standard quantum mechanics. Another
approach to calculate the arrival-time distribution is presented in
Section \ref{calc} and compared to the straight-forward calculation.
The advantage of this second approach is that is easily extended to
the full model, the corresponding calculations shown in
\ref{full_model}, and that it allows to some extent for an analytical
treatment of the arrival-time problem. In Section \ref{relation} we
discuss the relation of the present detection scheme to the
fluorescence model. Section \ref{discussion} deals with the
limit of zero coupling and an infinity of spins, and remarks on
possible schemes for the optimization of the model and on its
application to passage-time measurements.

\section{The detector model}
\label{model}

The detector model of \cite{lss1997, gs1990, lss1991} is based
on the following Hamiltonian. The excited state of the $j^\mathrm{th}$
spin is denoted by $\excited_j$ and its ground state by
$\ground_j$. Define
\begin{equation}
\hat{\sigma}_z^{(j)} \equiv
\excited_{j\;j}\!\excitedbra \, - \, \ground_{j\;j} \! \groundbra \,.
\label{detector}
\end{equation} 
The Hamiltonian for the detector alone is given by
\begin{equation} H_\mathrm{det} = \frac{1}{2}\sum_j \hbar\omega_0^{(j)}
\hat{\sigma}_z^{(j)} - \frac{1}{2}\sum_{j<k} \hbar\omega_J^{(jk)}
\hat{\sigma}_z^{(j)} \otimes \hat{\sigma}_z^{(k)}\,,
\label{hdet}
\end{equation} where $\hbar \omega_0^{(j)}$ is the energy difference
between ground state and excited state of the $j^\mathrm{th}$ spin,
and $\hbar\omega_J^{(jk)}\ge 0$ is the coupling energy between the
spins $j$ and $k$.

In addition there is a bath of bosons (e.g.\ phonons or photons) with
free Hamiltonian
\begin{equation}
H_\mathrm{bath} = \sum_{\bell} \hbar
  \omega_\ell \hat{a}_{\bell}^\dagger \hat{a}_{\bell},
\end{equation}
where $\hat{a}_{\bell}$ is the annihilation operator for a boson
with wave vector $\bell$. Later a continuum limit will be taken.
In general, the spins will be coupled to the bath, and there is the
possibility of spontaneous spin flips due to
\begin{equation}
  H_\mathrm{spon} = \sum_{j,\bell}\hbar \left(
    \gamma_{\bell}^{(j)} \rme^{\rmi f_{\bell}^{(j)}}
    \hat{a}_{\bell}^\dagger \hat{\sigma}_-^{(j)} + \mathrm{h.c.}
  \right),
\label{spont}
\end{equation}
where
\begin{equation}
\hat{\sigma}_-^{(j)} \equiv \ground_{j\;j} \!
\excitedbra, \quad \hat{\sigma}_+^{(j)} \equiv \left(
\hat{\sigma}_-^{(j)} \right)^\dagger = \excited_{j\;j} \! \groundbra,
\end{equation}
and the coupling constants $\gamma_{\bell}^{(j)}$ and the phases
$f_{\bell}^{(j)}$ depend on the particular realization of the
detector and the bath.

The coupling between the $j^\mathrm{th}$ spin and the bath is assumed
to be strongly enhanced when the particle is close to this spin. Let
the $j^\mathrm{th}$ spin be located in a spatial region ${\mathcal
  G}_j$.  The enhancement is taken to be proportional to 
a sensitivity function $\chi^{(j)} \left( \mathbf{x} \right)$ which
vanishes outside ${\mathcal G}_j$, e.g.\ the characteristic function
which is 1 on ${\mathcal G}_j$ and zero outside.
The additional coupling depending on
the particle's position is thus
\begin{equation}
  H_\mathrm{coup} = \sum_j \chi^{(j)} \left(
    \hat{\mathbf{x}} \right) \sum_{\bell}\hbar \left(
    g_{\bell}^{(j)} \rme^{\rmi f_{\bell}^{(j)}}
    \hat{a}_{\bell}^\dagger \hat{\sigma}_-^{(j)} + \hc \right),
\label{general_coupling}
\end{equation}
with $\left| g_{\bell}^{(j)} \right|^2 \gg \left|
  \gamma_{\bell}^{(j)} \right|^2$.

The full Hamiltonian is
\begin{equation} H = H_\mathrm{part} + H_\mathrm{det} +
H_\mathrm{bath} + H_\mathrm{spon} + H_\mathrm{coup} \,,
\label{hamiltonian}
\end{equation} where $H_\mathrm{part}$ is the free Hamiltonian of the
particle,
\begin{equation}\label{free} H_\mathrm{part} =
\hat{\mathbf{p}}^2/2m\,.
\end{equation}
Note that the `excitation number', i.e., the sum of the number of
bosons and the number of up-spins, is a conserved quantity. The
detection process now starts with the bath in its ground state
$\ket{0}$ (no bosons present) and all $D$ spins in the excited state
$\excitedall$. As a consequence of the excitation number conservation,
it is sufficient to measure the state $\ket{0}$ of the bath in order
to check whether or not any spin has flipped.  For $\hbar \omega_0^{(j)}$
only slightly above the energetic threshold set by the ferromagnetic
spin-spin coupling, and $\gamma_{\bell}^{(j)}$ sufficiently small,
the probability of a spontaneous spin flip (`false positive') is very
small \cite{lss1997, gs1990, lss1991}. But when the particle is close
to the $j^\mathrm{th}$ spin, the excited state $\excited_j$ decays
much more quickly, due to the enhanced coupling,
`$g_{\bell}^{(j)}$', of the spin to the bath.  Then, the
ferromagnetic force experienced by its neighbors is strongly reduced,
and thus these spins can flip rather quickly even in the absence of
the particle by means of the $\gamma_{\bell}^{(j)}$; by a kind of
'domino effect', the whole array of spins will eventually flip,
amplifying the first spin flip to a macroscopic event \cite{lss1997,
  gs1990, lss1991}.

\section{The direct approach in the one-spin case}
\label{direct_approach}

\subsection{A simplified model}

We first consider a simplified model consisting of a particle in one
dimension and only one spin. This simplification is reasonable if the
radius of the region ${\mathcal G}_j$ is smaller than the distance
between spins. (Our assumption of locality of the interaction though
is a bit stronger than this however, since below in Section
\ref{eigenstates}, for calculational convenience we will extend the
region ${\mathcal G}_j$ to a half-line, i.e., $\chi(x) \to \Theta
(x)$.)  The vectors $\mathbf{x}$ and $\bell$ are replaced by $x$ and
$\ell$.  Also, we will temporarily neglect $H_\mathrm{spon}$ in view
of assumption $\left| \gamma_\ell^{(j)} \right|^2 \ll \left|
  g_\ell^{(j)} \right|^2$, and accordingly the possibility of
spontaneous spin flips.

The free Hamiltonian for the particle motion in one dimension is
\begin{equation}
H^\one_\mathrm{part} = \hat{p}^2/2m,
\label{part_1d}
\end{equation}
and the free detector Hamiltonian with only one spin simplifies to
\begin{equation}
H^1_\mathrm{det} = \frac{1}{2}\hbar \omega_0 \hat{\sigma}_z.
\end{equation}
 The free bath
Hamiltonian is given by
\begin{equation}
H_\mathrm{bath}^\one = \sum_\ell \hbar \omega_\ell \hat{a}_\ell^\dagger
\hat{a}_\ell.
\label{bath_1d}
\end{equation}
Furthermore, let the spin be located in the interval $\IL \equiv [0,d]$ so
that
\begin{equation}
H^{1,\one}_\mathrm{coup} = \chi_\ILsub \left( \hat{x} \right)
\sum_\ell \hbar \left( g_\ell \rme^{\rmi f_\ell} \hat{a}_\ell^\dagger
\hat{\sigma}_- + \hc \right).
\label{general_coupling_one}
\end{equation}
where the sensitivity function $\chi_\ILsub( x)$ 
vanishes outside $\IL$. The full Hamiltonian of the simplified model
is then given by 
\begin{equation}
H^{1,\one} = H^\one_\mathrm{part} + H^1_\mathrm{det} +
H^\one_\mathrm{bath} + H^{1,\one}_\mathrm{coup}.
\label{hamiltonian_one}
\end{equation}
This simplified model allows for a direct investigation by means of
standard quantum mechanics.

\subsection{Energy eigenstates}
\label{eigenstates}

To get a first idea of how the present detector model works for an
arrival-time measurement, we simplify the model in this section a
little further by assuming the detector to be semi-infinite, extended
over the whole positive axis, and take momentarily 
 $$
\chi_\ILsub (x) = \Theta (x),
$$
where $\Theta$ is Heaviside's step function. Also, we assume for the
phases in the 
coupling Hamiltonian $f_\ell \equiv 0$ throughout this section. The
stationary Schr\"odinger equation with energy eigenvalue $E_k$ for a
plane wave coming in from the left, initially no bosons present, and the
spin in state $\ket{\uparrow}$, can be solved piecewise in position
space. For $x<0$, the solution simply reads
\begin{equation}
\boldsymbol{\Phi}^<_k (x) =\sqrt{\frac{1}{2\pi}} \left( \left[
    \rme^{\rmi kx} + R_0(k)
    \rme^{-\rmi kx} \right] \ket{\uparrow~0} 
+ \sum_\ell R_\ell (k) \rme^{-\rmi k_\ell(k)x} \ket{\downarrow~1_\ell} \right),
\end{equation}
where the wave numbers $k,~k_\ell(k)$ are fixed by
\begin{equation}
\frac{\hbar^2 k^2}{2m} + \frac{\hbar \omega_0}{2} = E_k = \frac{\hbar^2
k_\ell(k)^2}{2m} - \frac{\hbar \omega_0}{2} + \hbar \omega_\ell,
\label{energy_conservation}
\end{equation}
and where $\ket{\uparrow~0} \equiv \ket{\uparrow} \ket{0}$,
$\ket{\downarrow~1_\ell}\equiv \ket{\downarrow} \ket{1_\ell}$. Note that
there is the possibility 
that the particle is reflected from the detector. It may either be
reflected after it has been detected and a boson of mode $\ell$ has been
created, the coefficient for this event being $R_\ell(k)$, or it may even
be reflected without being detected, the coefficient being
$R_0(k)$. The latter will lead to a non-normalized arrival-time
distribution. Since this no-detection probability is in general
momentum dependent the momentum distribution of the actually detected
part of the wave packet must be expected to differ from that of the
originally prepared wave packet, hence leading to deviations of the
`measured' arrival-time distribution from corresponding `ideal'
quantities.

For $x>0$, the operator $H^{1,\one} -  \hat{p}^2/2m$ is independent of
$x$, because $\chi_\ILsub (x) = \Theta (x)$ has been assumed, and it
commutes with $\hat{p}^2/2m$. The
eigenvalues of $H^{1,\one} -  \hat{p}^2/2m$ are real and denoted
by $\hbar\Omega_\mu/2$. The corresponding eigenvectors are  superpositions
of $\ket{\uparrow~0}$ and $\ket{\downarrow~1_\ell}$ and denoted by
$\ket{\boldsymbol\mu}$ so that 
\begin{equation}
\left(H^{1,\one} -  \hat{p}^2/2m\right) \ket{\boldsymbol\mu} = \frac{\hbar
  \Omega_\mu}{2} \ket{\boldsymbol\mu}.
\end{equation}
To obtain an eigenvector of $H^{1,\one}$ on $x>0$ for the eigenvalue
$E_k$, one has to choose an eigenfunction $\rme^{\rmi q_\mu(k)x}$ of
$\hat{p}^2/2m$ such that
\begin{equation}
E_k = (\hbar q_\mu(k) )^2/2m + \hbar\Omega_\mu/2.
\end{equation}
From (\ref{energy_conservation}) one has
\begin{equation}
q_\mu(k) = \sqrt{ k^2 + \frac{m}{\hbar} \left( \omega_0 - \Omega_\mu
\right)}.
\end{equation}
Note that $q_\mu (k)$ is imaginary if $\Omega_\mu > \omega_0$ and
\begin{equation}
k^2 < \frac{m}{\hbar} \left( \Omega_\mu - \omega_0 \right),
\end{equation}
leading to exponential decay. Otherwise  $q_\mu(k)$ is real.
The  solution of the stationary Schr\"odinger equation  for $x>0$
belonging to the eigenvalue $E_k$  can then be written as
\begin{equation}
\boldsymbol{\Phi}_k^> (x) = \sqrt{\frac{1}{2\pi}} \sum_\mu \alpha_\mu(k)
\rme^{\rmi q_\mu(k)x} \ket{\boldsymbol\mu}.
\end{equation}
 The
coefficients $\alpha_\mu(k),~R_0(k),~R_\ell(k)$  are obtained from
the usual matching condition, i.e., both
\begin{equation}
\boldsymbol{\Phi}_k(x) := \cases{\begin{array}{ccc}
    \boldsymbol{\Phi}_k^< (x) & 
\mathrm{if} & x<0 \\ \boldsymbol{\Phi}_k^> (x) & \mathrm{if} & x \geq
0 \end{array}}
\end{equation}
and its first derivative have to be continuous at $x=0$. The
eigenvectors $\ket{\boldsymbol \mu}$ can be determined numerically.

\subsection{Detection of a wave packet}

The probability of finding the detector spin in state $\ket{\downarrow}$ (and
hence the bath in some boson state $|1_\ell\rangle$) at time $t$ is given
by integration over 
the modulus square of the respective component of $\ket{\boldsymbol{\Psi}_t}$,
\begin{eqnarray}
P_1^\mathrm{disc} (t) &=& \sum_\ell\int_{-\infty}^\infty \rmd x \, \left|
  \left\langle x~\downarrow~1_\ell \left|\boldsymbol{\Psi}_t
    \right. \right\rangle \right|^2\\
&=& 1 - \int_{-\infty}^\infty \rmd x \, \left|
  \left\langle x~\uparrow~0 \left|\boldsymbol{\Psi}_t \right. \right\rangle
\right|^2 \equiv 1 -P_0^\mathrm{disc}{}(t), \nonumber
\end{eqnarray}
where the superscript `$\mathrm{disc}$' distinguishes the
discrete model from the continuum limit discussed in the next section.
As long as no recurrences occur, i.e., no transitions
$\ket{\downarrow~1_\ell} \mapsto \ket{\uparrow~0}$, one can regard 
\begin{equation}
w_1^\mathrm{disc} (t) = \frac{\rmd}{\rmd t} P_1^\mathrm{disc}(t) = -
\frac{\rmd}{\rmd t} P_0^\mathrm{disc}(t). 
\end{equation}
as the probability density for a spin flip (i.e. for a
detection) at time $t$.

\begin{figure}[ht]
\begin{center}
\epsfig{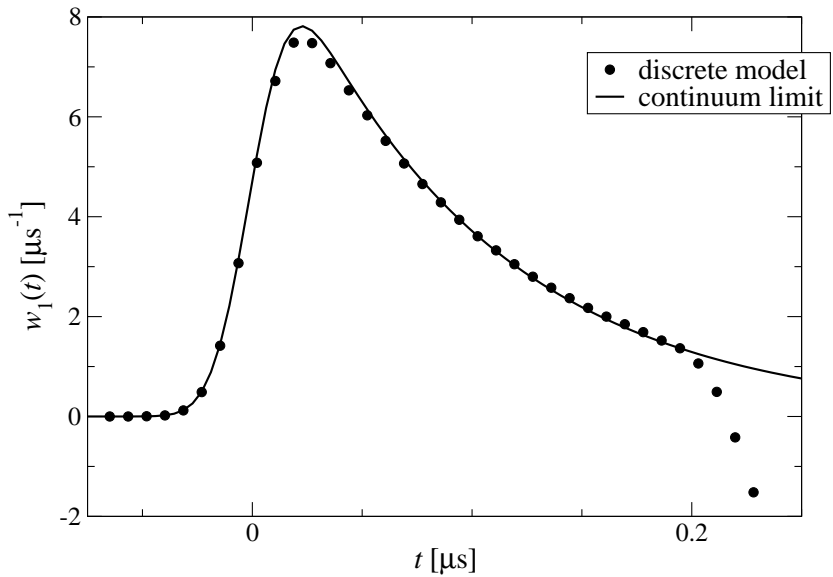}
\caption{Dots: spin-flip probability density $w_1^\mathrm{disc}(t)$
  for an incoming Gaussian wave packet of (\ref{state}) and
  (\ref{Gauss}) with $\Delta p= 20 \mu \mathrm{m}^{-1} \cdot \hbar$
  and $v_0= 1.79 \mathrm{m/s}$; $\omega_0 = 2.39 \cdot 10^8
  \mathrm{s}^{-1}$, $\omega_{_\mathrm{M}}=4.6 \cdot \omega_0$,
  $G=2.782 \cdot 10^3 \mathrm{s}^{-1/2}$, $N = 40$. Solid line:
  $w_1(t)$ from (\ref{w1-1d}) for the corresponding continuum limit.
  Up to the time of recurrences, $\ket{\downarrow~1_\ell} \mapsto
  \ket{\uparrow~0}$ (due to the discrete nature of the bath), the
  discrete and continuum probability densities are in good agreement.}
\label{w1discrete}
\end{center}
\end{figure}

As an example we consider a maximal boson frequency  $\omega_{_\mathrm{M}}$ 
and 
\begin{eqnarray}\label{ex}
\omega_\ell &=& \omega_{_\mathrm{M}} n/N ~~~~~n=1,\nonumber \cdots, N \\
g_\ell &=& -\rmi G \sqrt{\omega_\ell/N}. 
\end{eqnarray}
As particle we consider a cesium atom, prepared in the remote past far
away from the detector such that the corresponding free packet (i.e.,
in the absence of the detector) at $t=0$ would be a Gaussian minimal
uncertainty packet around $x=0$ with $\Delta p$ and average velocity
$v_0$. Decomposing this into the eigenstates of $H$, the wave packet
at time $t$ is
\begin{equation}\label{state}
\left\langle x \left| \boldsymbol{\Psi}_t \right. \right\rangle =
\int_{-\infty}^{\infty} \rmd k \,
\widetilde{\psi}(k) \boldsymbol{\Phi}_k(x) \rme^{-\rmi E_k t/\hbar}
\end{equation}
with
\begin{equation}\label{Gauss}
\widetilde{\psi} (k) = \left( \frac{\hbar }{\Delta p \sqrt{2 \pi}}
\right)^{1/2}  \exp \left(-\frac{\hbar^2}{4 (\Delta p)^2} \left( k -
  mv_0/\hbar \right)^2 \right).
\end{equation}
A numerical illustration of $w_1^\mathrm{disc}(t)$ for $N=40$ is 
given in figure \ref{w1discrete} (dots). The numerical calculation is
 time consuming, while in the continuous case with the quantum jump
approach it is much faster (see next section).

\section{Continuum limit and quantum jump approach}
\label{calc}

\subsection{Basic ideas}

In this section the detector model  and  its application to
arrival-times will be investigated in a continuum limit by 
means of the quantum jump approach \cite{qujump}. This approach uses
 continuous bath modes as a limit, so 
that there are no recurrences as in the discrete case. It is
easily generalized to multiple spins and it is more accessible to
analytic treatment. The bath modes are eliminated, but in contrast to
Bloch equations one can work with a (conditional or effective)
Hamiltonian and has reduced dimensions. It is based on watching for
the {\em first} appearance of a boson. To do this one would have to
observe the bath continuously. Since in
standard quantum mechanics with the simple von Neumann measurement
theory this would lead to difficulties associated with 
the quantum Zeno effect
\cite{beskow, khalfin, Sudarshan}, the quantum jump approach
 circumvents this by temporally coarse-grained observations and
a coarse-grained time scale. In the present situation, it reads as
follows. Instead of continuous observation, one considers repeated
instantaneous measurements, separated by a time $\Delta t$. For a
Markovian system with correlation time $\tau_\mathrm{c}$, one takes $\Delta t
\gg \tau_\mathrm{c}$ to avoid the quantum Zeno effect, but $\Delta t$ much shorter
than the lifetime of the excited state $\excitedall$ in order to
obtain a good time resolution. Typical numbers for quantum optical
models are $\Delta t \simeq 10^{-13} \mathrm{s} \ldots 10^{-10}
\mathrm{s}$. To find {\em no boson until $t = n \Delta t$}, no boson
must have been found in the first $n$ measurements. The probability
for this to happen will now be calculated. The detector interval $\IL$
can now be finite or semi-infinite.

Let the complete system (bath, detector, and particle) at $t_0=0$ be
prepared in the state
\begin{equation}
\ket{\Psi_0} = \nodetect \excitedall \ket{\psi_0},
\end{equation}
where $\ket{\psi_0}$ denotes the spatial wave function of the
particle. If no boson is found at the first measurement then, by the
von Neumann-L\"uders reduction rule \cite{vN,Lueders}, the state (up
to normalization) right after the measurement is given by projecting
with $\nodetectproj$,
\begin{equation} 
\ket{\Psi_\mathrm{cond}^{\Delta t}} \equiv
\nodetectproj U(\Delta t, 0) \nodetect \excitedall
\ket{\psi_0},
\label{state1}
\end{equation}
where $U(t,t')$ denotes the time evolution operator of
the complete system. The probability, $P_0(\Delta
t)$, for no detection is the norm squared of the vector in
(\ref{state1}), i.e. 
\begin{equation}
P_0(\Delta t)= \parallel \nodetectproj~ U(\Delta t,
0) \nodetect \excitedall \ket{\psi_0}\parallel ^2.
\end{equation}
The state then evolves with $U(2\Delta t, \Delta t)$ until the next
measurement, and so on. The state after the $n^\mathrm{th}$
consecutive no-boson measurement, $\ket{\Psi_\mathrm{cond}^{n \Delta
    t}}$, is, up to normalization,
\begin{eqnarray}
\ket{\Psi_\mathrm{cond}^{n \Delta t}} & = &
\nodetectproj U(n \Delta t, [n-1] \Delta t) \nodetect \cdots
\nonumber \\
& & \quad \quad \cdots
\bra{0} U(\Delta t, 0) \nodetect \excitedall
\ket{\psi_0}.
\label{n}
\end{eqnarray}
The probability, $P_0(n \Delta t)$, of finding
the bath in the state $\nodetect$ in \emph{all} of the first $n$
measurements is  given by its norm squared,
\begin{equation}
P_0(n \Delta t)= \left\langle
\Psi_\mathrm{cond}^{n \Delta t} \left| \Psi_\mathrm{cond}^{n \Delta t}
\right. \right\rangle.
\end{equation}
Note that $\bra{0} U(\nu\Delta t, [\nu-1] \Delta t)
\nodetect$ is an operator in the particle-detector Hilbert space which
does not rotate $\excitedall$, by excitation number conservation
mentioned after (\ref{free}). Thus one can write
\begin{equation}
\ket{\Psi_\mathrm{cond}^{n \Delta
t}}=\ket{\Psi_\mathrm{cond}^{t}} \equiv \nodetect \excitedall
\ket{\psi_\mathrm{cond}^t},
\end{equation}
where $t = n \Delta t$, and hence
\begin{equation}
\label{P0}
P_0(t) \equiv \left\langle
\Psi_\mathrm{cond}^{n \Delta t} \right| \left. \!
\Psi_\mathrm{cond}^{n \Delta t} \right\rangle = \left\langle
\psi_\mathrm{cond}^t
\right| \left. \! \psi_\mathrm{cond}^t \right\rangle
\end{equation}
is the
probability that no transition $\nodetect \longrightarrow \ket{1_\ell}$,
i.e.\ that no detection occurs \emph{until} the time $t$. The
probability  for the first detection to occur at next measurement
 is just given by
\begin{equation}\label{w0} 
P_0(t)- P_0(t+\Delta t)\equiv w_1(t) \Delta t.
\end{equation}
The crucial point now is to calculate the `conditional time
evolution' of $\ket{\psi_\mathrm{cond}^t}$, i.e., the
time evolution `under the condition that no detection occurs', and
for this one has to evaluate $\bra{0} U(\nu \Delta t, [\nu-1] \Delta t)
\nodetect\excitedall$. 

\subsection{A simplified model}

For greater clarity, the evaluation of $\bra{0} U(\nu \Delta t,
[\nu-1] \Delta t) \nodetect\excitedall$ will first be done for the
simplified model introduced at the beginning of Section
\ref{direct_approach}, while the generalization to the full model is
referred to \ref{full_model}.

We use the interaction picture w.r.t.\ $H_0^{1,\one} = H^{1,\one} -
H_\mathrm{coup}^{1,\one}$ and
 $U_I(t,t') = \rme^{\idh H_0^{1,\one} t} U(t,t') \rme^{- \idh
  H^{1,\one}_0 t'}$. Using (\ref{general_coupling_one}) with still
discrete, but possibly infinitely many, modes a simple calculation 
 gives in second order perturbation theory 
\begin{eqnarray}
\fl
\bra{0} U_I (\nu \Delta t, [\nu-1] \Delta t)
\nodetect \ket{\uparrow} = \ket{\uparrow} \left( \openone -
\phantom{\int\limits_{[\nu-1] \Delta t}^{\nu \Delta t} \rmd t_1}
\right. \nonumber \\
\left.
\int\limits_{[\nu-1] \Delta t}^{\nu \Delta t} \rmd t_1
\int\limits_{[\nu-1] \Delta t}^{t_1} \rmd t_2 \, \sum_\ell \chi_\ILsub \left(
\hat{x} \left( t_1 \right) \right)
\chi_\ILsub \left( \hat{x} \left( t_2 \right) \right)
\left| g_\ell \right|^2 \rme^{\rmi (\omega_0 - \omega_\ell) (t_1 - t_2)}
\right),
\label{Dt_full}
\end{eqnarray}
where $\hat{x} (t) = \hat{x} + \hat{p}t/m$ is the time
evolution of the operator $\hat{x}$ in the Heisenberg picture of the free
particle. The phases in the coupling terms have canceled; even if one
would assume these phases to be dependent on the particle's position,
$f_\ell(x)$, this would be the case to very good approximation since
$\Delta t$ is very small and thus $\hat{x} (t_1) \approx \hat{x}(t_2)$
\cite{gch2003}.
Consequently one obtains
\begin{eqnarray}
\fl
\bra{0} U_I (\nu \Delta t, [\nu-1] \Delta t)
\nodetect \ket{\uparrow} = \ket{\uparrow} \left( \openone - 
\phantom{\int\limits_{[\nu-1] \Delta t}^{\nu \Delta t} \rmd t_1}
\right. \nonumber \\
\left.
\int\limits_{[\nu-1] \Delta t}^{\nu \Delta t} \rmd t_1
\int\limits_{[\nu-1] \Delta t}^{t_1} \rmd t_2 \, \chi_\ILsub \left( \hat{x}
\left( t_1 \right) \right) \chi_\ILsub \left( \hat{x} \left( t_2 \right)
\right) \cdot \kappa \left( t_1 - t_2 \right) \right)
\label{Dt}
\end{eqnarray}
with the correlation function
\begin{equation} \kappa (\tau) \equiv \sum_\ell \left| g_\ell \right|^2 \rme^{-\rmi
(\omega_\ell - \omega_0) \tau}.
\end{equation} 

To have irreversible decay we go to the continuum
limit as follows. At first the bath modes are indexed by `wave numbers'
\begin{equation}\label{l}
\ell=2 \pi n/L_\mathrm{bath}, ~~~ n = 1,~2, \cdots
\end{equation}
and $\omega_\ell$ is chosen as
\begin{equation}\label{om}
\omega_\ell = c \left( \omega_\ell \right) \ell,
\end{equation}
so that
\begin{equation}\label{Delta}
\Delta \omega = \frac{c(\omega)^2}{c(\omega) - \omega c'(\omega)}
\Delta \ell = \frac{c(\omega)^2}{c(\omega) - \omega c'(\omega)} \cdot
\frac{2\pi}{L_\mathrm{bath}}.
\end{equation}
The coupling constants are taken to be of the form
\begin{equation} g_\ell =\left( \Gamma (\omega_\ell) +
    \mathcal{O}(L_\mathrm{bath}^{-1})\right) \cdot
  \sqrt{\frac{\omega_\ell}{L_\mathrm{bath}}} 
\label{coupling_constants}
\end{equation} where $\Gamma(\omega_\ell)$ does not depend on
$L_\mathrm{bath}$.
Then one obtains in the continuum limit by (\ref{Delta})
\begin{equation}\label{kappa}
\kappa (\tau) = \frac{1}{2\pi}\int_0^\infty \rmd\omega \, \frac{c(\omega)
  - \omega c'(\omega)}{c(\omega)^2} \omega |\Gamma
(\omega)|^2 \rme^{-\rmi(\omega - \omega_0) \tau}
\end{equation}
We assume  $\Gamma(\omega)$ to be of such a form that the Markov
property holds, i.e. 
\begin{equation} 
\kappa (\tau) \approx 0~~~ \mathrm{if} ~~~\tau > \tau_\mathrm{c}
\label{markovianproperty}
\end{equation} 
for some small correlation time $\tau_\mathrm{c}$. This is the case, e.g., for
$\Gamma(\omega) \equiv \Gamma$  as in quantum optics.
In the double integral of (\ref{Dt}) then only times with $t_1 - t_2 \le
\tau_\mathrm{c}$ contribute, and if   $\tau_\mathrm{c}$ is small enough one can write
\begin{equation} 
\chi_\ILsub \left( \hat{x} \left( t_1 \right) \right)
\chi_\ILsub \left( \hat{x} \left( t_2 \right) \right) \approx \chi_\ILsub
\left( \hat{x} \left( t_1 \right) \right)^2.
\label{approx}
\end{equation} 
The double integral then becomes
\begin{equation}
\int_0^{\Delta t} \rmd t' \chi_\ILsub(\hat{x}(t' + (\nu-1)\Delta t))^2
\int_0^{t'} \rmd\tau \, \kappa(\tau). 
\label{double}
\end{equation}
With $\Delta t \gg
\tau_\mathrm{c}$ the second integral can be extended to infinity, by the Markov
property. Putting
\begin{eqnarray}
A &\equiv& 2 \mathrm{Re} \int_0^\infty \rmd\tau \, \kappa(\tau)
= \omega_0 \frac{c(\omega_0) - \omega_0 c'(\omega_0)}{c(\omega_0)^2}
\cdot \left| \Gamma  \left(\omega_0 \right) \right|^2 \\ \nonumber
\delta_\mathrm{shift}  &\equiv& 2 \mathrm{Im}  \int_0^\infty \rmd\tau \,
\kappa(\tau)
\label{A}
\end{eqnarray}
one obtains
\begin{eqnarray}
\fl
\bra{0} U_I(\nu \Delta t,  [\nu-1] \Delta t) \nodetect
\ket{\uparrow } & = & \ket{\uparrow} \left( \openone - \frac{1}{2}(A
  + \rmi\delta_\mathrm{shift})
\int\limits_{[\nu-1] \Delta t} ^{\nu \Delta t} \!\! \rmd t_1 \, \chi_\ILsub \!
\left( \hat{x} \left(t_1\right) \right)^2 \right) \nonumber \\
& = & \ket{\uparrow}
\exp \left( - \frac{1}{2}(A  + \rmi\delta_\mathrm{shift})
  \int\limits_{[\nu-1] \Delta t} ^{\nu \Delta t} \, \rmd t_1 \,
  \chi_\ILsub \left( \hat{x} \left(t_1\right) \right)^2 \right),
\end{eqnarray}
up to higher orders in $\Delta t$. Note that $A$ is a decay rate of
the upper spin level; in
quantum optics  $ A$ and $\delta_\mathrm{shift}$ correspond to the Einstein
coefficient and to a line shift. 
 Going back to the Schr\"odinger picture
 one then obtains by (\ref{n})
\begin{equation}\label{cond}
 \ket{\psi_\mathrm{cond}^t}= \rme^{- \idh
\Hcond (t-t_0)} \ket{\psi_0}
\end{equation}
with the `conditional Hamiltonian'
\begin{equation}
\Hcond \equiv \frac{\hat{p}^2}{2m} + 
\frac{\hbar}{2}(\delta_\mathrm{shift} - \rmi A)\chi_\ILsub (\hat{x})^2.
\end{equation}
Note that this result is independent of the particular choice of
$\Delta t$ as long as $\Delta t$  satisfies the above requirements.
As a consequence, on a coarse-grained time scale in which $\Delta t$ is
small, $t$ can be regarded as continuous and
$\ket{\psi_\mathrm{cond}^t}$ obeys a Schr\"odinger equation with a 
complex potential,
\begin{equation} \rmi\hbar \frac{\partial}{\partial t}
\ket{\psi_\mathrm{cond}^t} = \left( \frac{\hat{p}^2}{2m}
 +  \frac{\hbar}{2} (\delta_\mathrm{shift} - \rmi A)\chi_\ILsub (\hat{x})^2\right)
\ket{\psi_\mathrm{cond}^t}.
\label{absorbing_potential}
\end{equation}
In this continuous, coarse-grained, time scale, (\ref{w0}) yields
for the probability density, $w_1(t)$, for the first detection
\begin{equation}
w_1(t) = -\frac{\rmd P_0(t)}{\rmd t}
\label{w1}
\end{equation}
and from (\ref{P0}) and (\ref{cond})
one easily finds  
\begin{eqnarray}
w_1(t) & = & \frac{\rmi}{\hbar} \left\langle \psi_\mathrm{cond}^t
  \left|\Hcond - \Hcond^\dagger \right| \psi_\mathrm{cond}^t
\right\rangle \nonumber \\
& = & A \int_0^d \rmd x \,\chi_\ILsub (x)^2 \left| \left\langle
x \left| \psi_\mathrm{cond}^t \right. \right\rangle
\right|^2.
\label{w1-1d}
\end{eqnarray}
If $\chi_\ILsub (\hat{x})$ is the characteristic function of the interval
[0,$d$] this is just the decay rate of the excited state of the
detector multiplied by the probability that the particle is inside the
detector but not yet detected --- a very physical result.

\subsection{An example}

As an example we consider the continuum limit of the discrete model of
(\ref{ex}). In this case  one has, with 
 $\omega_{_\mathrm{M}}$ the maximal frequency,
\begin{eqnarray}\label{excont}
c(\omega) & \equiv & c_0 \nonumber \\
\omega_\ell &=& \omega_{_\mathrm{M}} n/N \equiv c_0 \, 2\pi n/L_\mathrm{bath},
~~~n=1, \cdots, N \nonumber\\
L_\mathrm{bath} &\equiv& 2\pi c_0 N/\omega_{_\mathrm{M}} \nonumber\\
g_\ell &=& -\rmi G \sqrt{\omega_\ell/N} \equiv -\rmi G \sqrt{2\pi
  c_0/\omega_{_\mathrm{M}}}\sqrt{\frac{\omega_\ell}{L_\mathrm{bath}}}\nonumber\\
\Gamma\left( \omega \right) &=& \cases{%
  \begin{array}{cc} -\rmi G \sqrt{2\pi c_0/\omega_{_\mathrm{M}}}\equiv
    \Gamma & \mathrm{if} \;\; \omega \le \omega_{_\mathrm{M}} \\ 0 &
    \mathrm{else}. \end{array}}
\end{eqnarray}
In the continuum limit, $N$ or $L_\mathrm{bath}\to \infty$, one
obtains in the case $\omega_{_\mathrm{M}} > \omega_0$
\begin{eqnarray}
\kappa (\tau) & = & \frac{|G|^2}{\omega_{_\mathrm{M}}} \cdot \frac{
  \left( 1 + \rmi \omega_{_\mathrm{M}} \tau \right) \rme^{-\rmi
    (\omega_{_\mathrm{M}}-\omega_0) \tau} -
  \rme^{\rmi \omega_0 \tau}}{\tau^2} \nonumber \\
A &=&  2\pi \left| G \right|^2
\,\frac{\omega_0}{\omega_{_\mathrm{M}}}\nonumber \\
\delta_\mathrm{shift}&=& 2 \left|G\right|^2
\left(\frac{\omega_0}{\omega_{_\mathrm{M}}} \ln
  \left[\frac{\omega_0}{\omega_{_\mathrm{M}} - \omega_0} \right] -
  1\right)
\end{eqnarray}
and $\tau_\mathrm{c}$ is of the order of $\omega_0^{-1}$.
In the integral for $w_1(t)$ in (\ref{w1-1d}) one has $\chi_\ILsub
(x) = \Theta (x)$. The resulting $w_1(t)$ is plotted in figure
\ref{w1discrete} for the same wave function and parameters as for
$w_1^\mathrm{disc}(t)$ in that figure. Both distributions are in good
agreement up to the occurrence of recurrences $\ket{\downarrow~1_\ell}
\mapsto \ket{\uparrow~0}$ in the discrete case.

The agreement is also seen for other values of $\Delta p$. If $\Delta
x$ is the width of the wave packet in position space and $v_0$ is its
average velocity, then the width of the probability density for
detection is at least of the order of $\Delta x / v_0$ since it takes
some time for the wave packet to enter the detection region.  (Further
broadening of the detection density arises from the width of the delay
of the first spin flip once the particle is inside the detector.)
Consequently, wave packets with small $\Delta p$ and thus large
$\Delta x$ yield rather broad detection densities.  On the other hand,
as soon as a significant part of the wave function overlaps the
detector, in the discrete case the time scale of the recurrences is
essentially determined by the properties of the detector and the bath
and by their coupling. In case of wave packets with small $\Delta p$
long recurrence times are needed to obtain a good resolution of the
typically broad detection densities. This requires a large number of
bath modes in the discrete case. We further note that for more
complicated incident wave packets as, e.g., the coherent superposition
of several Gaussian wave packets with different mean velocities, the
probability density exhibits a more complicated structure due to the
self-interference of the wave function.

\subsection{The general case}

A procedure analogous to (\ref{Dt_full}) - (\ref{w1-1d})
can be applied to the three-dimensional model  
with several spins, as explained in \ref{full_model}. The bosons are
allowed to have a direction $\mathbf{e}$ which varies over the unit
sphere. In a  continuum limit  $\ket{\psi_\mathrm{cond}^t}$ obeys a
Schr\"odinger equation with a complex potential
\begin{equation}
\rmi\hbar \frac{\partial}{\partial t}
\ket{\psi_\mathrm{cond}^t} = \Hcond \ket{\psi_\mathrm{cond}^t}
\label{three-d-schroedinger}
\end{equation}
where 
\begin{equation}\label{three-d-cond}
\Hcond =
\frac{\hat{\mathbf{p}}^2}{2m} + 
\frac{\hbar}{2}\{\delta_\mathrm{shift}(\hat{\mathbf{x}}) -\rmi
A(\hat{\mathbf{x}})\}.  
\end{equation}
$A(\mathbf{x})$ and $\delta_\mathrm{shift}({\mathbf{x}})$ are given in
(\ref{Athree}) and (\ref{3d-shift}). The probability density for
the first detection is again similar to (\ref{w1-1d}),
\begin{eqnarray}
w_1 (t) & = & \frac{\rmi}{\hbar} \left\langle \psi_\mathrm{cond}^t
  \left|\Hcond - \Hcond^\dagger \right| \psi_\mathrm{cond}^t
\right\rangle \nonumber\\
& = & \int \rmd^3 x \; A \left( \mathbf{x} \right) \left|
  \left\langle \mathbf{x} \left| \psi_\mathrm{cond}^t
    \right. \right\rangle \right|^2,
\label{w}
\end{eqnarray}
which is an average of the position dependent decay rate of the
detector, weighted with the probability density for the particle to be
at position $\mathbf{x}$ and yet undetected.

\section{Relation of the present detector model to the fluorescence
  model}
\label{relation}

In the quantum optical fluorescence model \cite{dambo2002, hsm2003,
  bn2003b, dambo2003, hsmn2004, rdnmh2004, hhm2005} for arrival times
one considers a two-level atom with ground state $\ket{1} \equiv {1
  \choose 0}$ and excited state $\ket{2} \equiv {0 \choose 1}$, which
enters a laser illuminated region. In the one-dimensional case one
obtains a conditional Hamiltonian of the form
\begin{equation} \Hcond^\mathrm{fl} = \frac{\hat{p}^2}{2m} +
\frac{\hbar}{2} \left(
  \begin{array}{cc} 0 & \Omega(\hat{x}) \\ \Omega (\hat{x}) & -\rmi
\gamma - 2\Delta \end{array} \right),
\end{equation} where $\Omega (x)$ is the (position dependent) Rabi
frequency of the laser, $\Delta$ the detuning (possibly also position
dependent), and $\gamma$ the decay constant of the excited level. Note
that in contrast to the present model this is a two-channel
Hamiltonian. The reason for this is that the ground state of the
quantized photon field is not related to a specific internal state of
the atom due to the driving by the (classical) laser. In the limit
\begin{equation} \frac{\hbar | 2 \Delta + \rmi \gamma|}{2} \gg
\frac{\hbar}{2} \Omega,~ E
\label{onechannel}
\end{equation} where $E$ denotes the kinetic energy of the incident
particle, the corresponding conditional Schr\"odinger equation reduces
to a one-channel equation for the ground state amplitude with the
complex potential
\begin{equation} V(x) = \frac{\hbar \Delta \Omega(x)^2 - \rmi \hbar
\gamma \Omega(x)^2 / 2}{4 \Delta^2 + \gamma^2},
\end{equation}
and the excited state can be neglected in this limit \cite{bn2003b}.
Physically, condition (\ref{onechannel}) means that the excited state
decays very rapidly compared to the time-scales of the pumping and the
center-of-mass motion. Thus, the first fluorescence photon is emitted,
i.e., the particle is detected, when and where the excitation takes
place. For $\Delta \equiv 0$ (laser in resonance) $V$ is a purely
imaginary potential, similar as for the detector model outlined above;
only the physical interpretation of the height of this imaginary
potential differs. In other words, the one-channel limit of the
fluorescence model coincides with the full quantum mechanical model
from Section \ref{model} when considering the conditional interaction
for the particle until the first detection. In this way, the fully
quantum mechanical detector model of Section \ref{model} not only
justifies the fluorescence model for quantum arrival times, at least
in the limit of (\ref{onechannel}), but one can conversely immediately
carry over the results of the fluorescence model to the detector
model. The investigation of the fluorescence model has shown that the
essential features like reflection and delay \cite{dambo2002}, and
main results like, e.g., linking Kijowski's arrival-time distribution
to a particular measuring process \cite{hsm2003}, can be obtained from
the full two-channel model as well as from its one-channel limit.
Hence these results immediately carry over to the present detector
model. Also, the derivation of a complex potential model for particle
detection from two different physical models, viz.\ the fluorescence
model and the present detector model, indicates the importance of the
complex potentials and of Kijowski's arrival-time distribution, which
in turn can be derived from the complex potentials approach. This
connection is interesting since it can illuminate the physical
background of otherwise heuristically introduced complex potentials.
Differences, however, arise for example in applications to passage
times since the reset state after a detection is not the same in the
two models \cite{nhs}.

\section{Discussion and extensions}
\label{discussion}

The model investigated in this paper has three ingredients, viz. a
particle in whose spatial properties one is interested, a `detector'
based on spins, and a bath of bosons, originally in the ground state.
There is neither a direct measurement on the particle of interest nor
on the detector but only on the bath, which is checked for bosons. In
this way one can hope to keep the disturbance of the particle by the
measurement to a minimum. However, as in the fluorescence model and
also seen in \cite{jjh1999} for a simplified spin model, also the
present full model yields a description by a complex potential and
thus shows the typical unwanted features: There is a detection delay,
due to the finite spin decay or flip rate, and there is also
necessarily the possibility of reflection of the particle by the
detector without the detection of a boson, due to the increased
bath-detector coupling caused by the particle's wave function inside
the detector. This reflection without boson detection causes a
non-detection of the particle so that the probability density $w_1(t)$
in (\ref{w}) for the first detection is not normalized. A similar
effect arises from the transmission of the particle without boson
detection.

In order to reduce the detection delay one may be tempted to increase
the spin-bath coupling, which mirrors the particle's wave function
inside the detector. As a by-product this would also decrease
transmission without detection. However, the increase of this
spatially dependent coupling means an increase of the absorbing
potential $-\rmi \hbar A(\mathbf{x})/2$, and this will also increase the
reflection without boson detection, so much so that in the limit of
infinite coupling everything is reflected while nothing is detected.
The same phenomenon occurs in the fluorescence model \cite{dambo2002}
and is a typical feature of complex potentials, as already noted by Allcock~\cite{ga1969}.

One can also try to reduce the influence of the spin-bath
system on the particle and thus the latter's disturbance by decreasing
the spin-bath coupling at a space point and simultaneously increasing
the number of spins located there. This seems natural because it is
the flip of a single spin which gives rise to the detection, and with
a larger number of spins this can compensate for the weaker
coupling. To investigate this quantitatively we consider $N$ spins,
later to be taken to $\infty$, in the same volume $V$ and
$\chi^{(j)}(\mathbf{x})\equiv \chi_V(\mathbf{x})$ for all $j$. The
coupling constants are taken in the form
\begin{equation}
g_{\bell}^{(j)} \equiv g_{\bell} =
\frac{\Gamma \left(\omega_{\bell}, \mathbf{e}_{\bell}
  \right) + {\mathcal O} \left(L_\mathrm{bath}^{-1} \right)}{\sqrt{N}}
\cdot \sqrt{\frac{\omega_\ell}{L_\mathrm{bath}^3}}
\end{equation}
 and similarly for
$\gamma_{\bell}^{(j)}$. Further, the ferromagnetic force
experienced by the individual spin is assumed not to grow with
increasing $N$ such as for nearest neighbor interaction. Then 
(\ref{Athree}) becomes
\begin{eqnarray}
\fl
A \left( \mathbf{x} \right) & = & \sum_{j=1}^N
\left( \tilde{\omega}_0 \right)^3
\left[\frac{c \left( \tilde{\omega}_0 \right) -
    \tilde{\omega}_0 c' \left( \tilde{\omega}_0 \right)}{c
    \left( \tilde{\omega}_0  \right)^4} \right] \int
\frac{\rmd \Omega_\mathbf{e}}{(2 \pi)^2} \, \frac{\left| \Gamma  \left(
\tilde{\omega}_0 , \mathbf{e} \right) \right|^2
\chi_V ({\mathbf{x}})^2 
 + \left| \Gamma _\mathrm{spon} \left(
\tilde{\omega}_0 , \mathbf{e} \right) \right|^2}{N}
 \nonumber \\
\fl
& = & \left( \tilde{\omega}_0 \right)^3
\left[\frac{c \left( \tilde{\omega}_0 \right) -
    \tilde{\omega}_0 c' \left( \tilde{\omega}_0 \right)}{c
    \left( \tilde{\omega}_0  \right)^4} \right] \int
\frac{\rmd \Omega_\mathbf{e}}{(2 \pi)^2} \left( 
\left| \Gamma  \left(
\tilde{\omega}_0 , \mathbf{e} \right) \right|^2
\chi_V ({\mathbf{x}})^2 
 + \left| \Gamma _\mathrm{spon} \left(
\tilde{\omega}_0 , \mathbf{e} \right) \right|^2 \right)
\end{eqnarray}
which is just the decay rate for a single spin in $V$,
with resonance frequency $\tilde{\omega}_0$ and the coupling as for
$N=1$. A similar result holds for $\delta_\mathrm{shift}
(\mathbf{x})$, defined in (\ref{3d-shift}).

Thus, simply increasing the number of spins $N$ and scaling the
coupling constants with $\sqrt{1/N}$ leaves $A$ and
$\delta_\mathrm{shift}$ invariant and thus does not change the
dynamics until the first detection (spin flip), and in particular does
not help to avoid reflection of undetected particles.  Any other
scaling power of $N$, however, would not lead to a reasonable detector
model in the limit $N \rightarrow \infty$ since then either $A$ and
$\delta_\mathrm{shift}$ would go to zero or to $\infty$. Similar
results also hold for the quantum optical fluorescence model. It is
interesting to note that, although it is the flip of one single spin
which triggers the detection, it is the totality of all spins located
in $V$ which determines the conditional time evolution.

It has been shown in the context of complex potentials, however, that
one can deal with the delay/transmission-versus-reflection problem by
dropping the restriction to rectangular potentials \cite{mbm1995,
  pms1998}. In fact, it is possible to absorb nearly the complete wave
packet in a very short spatial interval; given a wave packet
with a specific energy range, an appropriate imaginary potential
can be constructed by means of inverse scattering techniques. We
stress that there is no such a thing as \emph{the} optimal imaginary
potential for all wave packets but the construction of the optimized
potential requires \emph{a priori} information about energy range of
the wave packet under consideration.

The present detector model is applicable not only to arrival-time
measurements, but also to more involved tasks like a measurement of
passage times. A detailed analysis including numerical examples will
appear elsewhere \cite{nhs}. It turns out that a too weak spin-bath
coupling yields a broad passage-time distribution due to the slow
response of the detector to the presence of the particle. A too strong
spin-bath coupling, on the other hand, also yields a broad
passage-time distribution due to the strong distortion of the wave
packet during the measurement process. This is a quantum effect. There
is, however, an intermediate range for $A(\mathbf{x})$ yielding rather
narrow passage-time distributions. Indeed, a rough estimate in
\cite{nhs} shows that for an optimal choice of incident wave packet
and decay rate $A (\mathbf{x})$ the precision of the measurement can
be expected to behave like $E^{-3/4}$, where $E$ is the energy of the
incident particle.  For low velocities, this means some improvement
as compared to the results of models coupling the particle
continuously or semi-continuously to a clock, where one has
$E^{-1}$-behavior \cite{ap1980, amm2003}. Thus, it appears
that the latter $E^{-1}$ behavior of the precision is not due to a
fundamental limitation related to a kind of time-energy uncertainty
relation.

\section*{Acknowledgements}

This work was supported in part by NSF Grant PHY 0555313.

\section*{Summary}

We have investigated the continuum limit of a fully quantum
mechanical spin-model for the detection of a moving particle when
the spin-boson interaction satisfies the Markov property.
In an example with a single spin and 40 boson modes it was
shown numerically that the continuum limit gave a good approximation
to the discrete model up to times of revivals.  We have derived
analytical expressions for the arrival-time distribution. The
conditional Schr\"odinger equation governing the particle's time
evolution before the detection has the same form as the one-channel
limit of the fluorescence model, which is based on the use of a
laser-illuminated region. The quantum spin detector-model provides an
easier way to obtain this one-channel equation, since no additional
assumptions or limits are needed.

\appendix

\section{The quantum jump approach for several spins}
\label{full_model}

The continuum limit and quantum jump approach for the full model
 in Section \ref{model} is 
quite similar until (\ref{Dt_full}). In second order perturbation
theory w.r.t.\ $H_\mathrm{coup} + H_\mathrm{spon}$ one obtains
\begin{eqnarray}
\fl
  \bra{0} U_I (\nu \Delta t, [\nu-1] \Delta t) \ket{0} \excitedall =
  \excitedall
\left( \openone - \sum_{j,\bell} \int\limits_{[\nu-1] \Delta t}^{\nu
      \Delta t} \rmd t_1 \int\limits_{[\nu-1] \Delta t}^{t_1} \rmd t_2 \,
\right. \nonumber \\
\fl \left. \phantom{\int\limits_{[\nu-1] \Delta t}^{\nu \Delta t} \rmd
    t_1}
  \rme^{\rmi
      \left( \widetilde{\omega}_0^{(j)} -
        \omega_\ell \right) \left( t_1 - t_2 \right)}
    \overline{\left(
        \chi^{(j)} \left( \hat{\mathbf{x}} \left( t_1 \right) \right)
        g_{\bell}^{(j)} + \gamma_{\bell}^{(j)} \right)} \cdot
    \left( \chi^{(j)} \left( \hat{\mathbf{x}} \left( t_2 \right)
      \right) g_{\bell}^{(j)} + \gamma_{\bell}^{(j)} \right)
  \right)
\label{Dt_mult_three}
\end{eqnarray}
where
\begin{equation}
\widetilde{\omega}_0^{(j)} \equiv \omega_0^{(j)} -
  \left(\sum_{k=1}^{j-1} \omega_J^{(kj)} + \sum_{k=j+1}^D
    \omega_J^{(jk)} \right)
\end{equation}
are modified resonance frequencies arising from the ferromagnetic
spin-spin coupling.  The phases $f_{\bell}^{(j)}$ have canceled
similar to the one-spin case since only products of the form
$\hat{a}_\ell \hat{\sigma}_+^{(j)} \hat{a}_\ell^\dagger
\hat{\sigma}_-^{(j)}$ contribute to the second order, and consequently
the contributions from different spins do not mix.

Similar to (\ref{kappa}) one can define correlation functions
$\kappa^{(j)}_{\overline{g}g}$, $\kappa^{(j)}_{\overline{g}\gamma}$,
$\kappa^{(j)}_{\overline{\gamma}g}$ and
$\kappa^{(j)}_{\overline{\gamma}\gamma}$ in an obvious way. Before the
continuum limit the bath modes are indexed by the wave vectors
\begin{equation}
\bell = \frac{2 \pi}{L_\mathrm{bath}} \left(
\begin{array}{c} n_1 \\ n_2 \\ n_3 \end{array} \right), \;\; n_i =1,~2,~\cdots
\end{equation}
In analogy
to (\ref{coupling_constants}), the coupling constants are taken
in the form
\begin{eqnarray}\label{coupling_constants_three}
 g_{\bell}^{(j)}&=& \left(\Gamma^{(j)} (\omega_\ell,
   \mathbf{e}_{\bell}) + \mathcal{O}(L_\mathrm{bath}^{-1})\right) \cdot 
\sqrt{\frac{\omega_\ell}{L_\mathrm{bath}^3}}\\
 \gamma_{\bell}^{(j)} &=&
\left(\Gamma_\mathrm{spon}^{(j)} (\omega_\ell, \mathbf{e}_{\bell}) +
  \mathcal{O}(L_\mathrm{bath}^{-1})\right) \cdot 
\sqrt{\frac{\omega_\ell}{L_\mathrm{bath}^3}}
\end{eqnarray}
with $\omega_\ell = c (\omega_\ell) \ell$, $\mathbf{e}_{\bell} =
\bell/\ell$, and
\begin{equation}
\label{gg} 
\left| \Gamma^{(j)} (\omega_\ell, \mathbf{e}_{\bell}) \right|^2 \gg
\left| \Gamma_\mathrm{spon}^{(j)} (\omega_\ell,
  \mathbf{e}_{\bell}) \right|^2.
\end{equation}
Again the Markov property is assumed to hold for the correlation
functions in the continuum limit. The procedure is then analogous to
the single spin case, and one obtains (\ref{cond}) with the
conditional Hamiltonian
\begin{equation}
H_\mathrm{cond} = 
\frac{\hat{\mathbf{p}}^2}{2m} + \frac{\hbar}{2} \{\delta_\mathrm{shift}
 (\hat{\mathbf{x}} ) - \rmi A(\hat{\mathbf{x}} ) \} 
\end{equation}
where $A(\mathbf{x})$  is given  in analogy to (\ref{A}) by
\begin{eqnarray}\label{Athree}
 A\left(\mathbf{x} \right) &=& 2\, \mathrm{Re}\, \sum_j
 \int_0^\infty \rmd\tau \,
\{\kappa_{\overline{g}g}^{(j)}(\tau) \chi^{(j)}({\mathbf{x}})^2 +
\kappa_{\overline{\gamma}\gamma}^{(j)}(\tau)\} \\ 
&=& \sum_j \left( \tilde{\omega}_0^{(j)} \right)^3
\left[\frac{c \left( \tilde{\omega}_0^{(j)} \right) -
    \tilde{\omega}_0^{(j)} c' \left( \tilde{\omega}_0^{(j)} \right)}{c
    \left( \tilde{\omega}_0^{(j)} \right)^4} \right] \int
\frac{\rmd \Omega_\mathbf{e}}{(2 \pi)^2}\nonumber\\
&& 
\phantom{\sum_j \left( \tilde{\omega}_0^{(j)} \right)^3} \times
\left( 
\left| \Gamma^{(j)} \left(
\tilde{\omega}_0^{(j)}, \mathbf{e} \right) \right|^2
\chi^{(j)}({\mathbf{x}})^2 
 + \left| \Gamma^{(j)}_\mathrm{spon} \left(
\tilde{\omega}_0^{(j)}, \mathbf{e} \right) \right|^2 \right)\nonumber
\end{eqnarray}
where the $\rmd\Omega_\mathbf{e}$ integral is taken over the unit sphere
and where the contributions from $\kappa^{(j)}_{\overline{g}\gamma}$,
$\kappa^{(j)}_{\overline{\gamma}g}$ have been neglected, due to
(\ref{gg}). The terms have the familiar form of the Einstein
coefficients in quantum optics, where there would also be a sum over
polarizations.
 $\delta_\mathrm{shift} \left(\mathbf{x}\right)$ is given by
\begin{equation}\label{3d-shift}
  \delta_\mathrm{shift} \left(\mathbf{x}\right) = 
  2\, \mathrm{Im}\, \sum_j \int_0^\infty \rmd\tau
  \{\kappa_{\overline{g}g}^{(j)}(\tau) \chi^{(j)}({\mathbf{x}})^2 +
  \kappa_{\overline{\gamma}\gamma}^{(j)}(\tau)\}.
\end{equation}
Since the $\kappa_{\overline{\gamma}\gamma}$ term leads to a constant it just
gives an overall phase factor and can therefore be omitted.



\section*{References}

\begin{thebibliography}{10}

\bibitem{sgad1996}
Szriftgiser P, Gu\'ery-Odelin D, Arndt M and Dalibard J 1996
\newblock {\em Phys. Rev. Lett.} {\bf 77} 4

\bibitem{mme2002}
Muga J~G, {Sala Mayato} R and Egusquiza I~L (eds) 2002
\newblock {\em Lecture Notes
  in Physics} vol m 72 {\em Time in Quantum Mechanics} 
\newblock (Berlin Heidelberg: Springer Verlag)

\bibitem{lss1997}
Schulman L~S 1997
\newblock {\em Time's Arrows and Quantum Mea\-sure\-ment}
\newblock (Cambridge: Cambridge University Press)

\bibitem{ga1969}
Allcock G~R 1969
\newblock {\em Ann. of Phys. (NY)} {\bf 53} 253, 286, 311

\bibitem{jk1974}
Kijowski J 1974
\newblock {\em Rep. Math. Phys.} {\bf 6} 361

\bibitem{rg1997}
Giannitrapani R 1997
\newblock {\em Int. J. Theor. Phys.} {\bf 36} 1575

\bibitem{ab_toa1961}
Aharonov Y and Bohm D 1961
\newblock {\em Phys. Rev.} {\bf 122} 1649

\bibitem{oru1999}
Oppenheimer J, Reznik B and Unruh W.~G 1999
\newblock {\em Phys. Rev.} A {\bf 59} 1804

\bibitem{crl2002}
Leavens C~R 2002
\newblock {\em Phys. Lett.} A {\bf 303} 154

\bibitem{emnr2003}
Egusquiza I~L, Muga J~G, Navarro B and Rusch\-haupt A 2003
\newblock {\em Phys. Lett.} A {\bf 313} 498

\bibitem{crl2005}
Leavens C~R 2005
\newblock {\em Phys. Lett.} A {\bf 345} 251

\bibitem{jjh1999}
Halliwell J~J 1999
\newblock {\em Progr. Theor. Phys.} {\bf 102} 707

\bibitem{gs1990}
Gaveau B and Schulman L~S 1990
\newblock {\em J. Stat. Phys.} {\bf 58} 1209

\bibitem{lss1991}
Schulman L~S 1991
\newblock {\em Ann. Phys. (NY)} {\bf 212} 315

\bibitem{dambo2002}
Damborenea J~A, Egusquiza I~L, Hegerfeldt G~C and Muga J~G 2002
\newblock {\em Phys. Rev.} A {\bf 66} 052104

\bibitem{hsm2003}
Hegerfeldt G~C, Seidel D and Muga J~G 2003
\newblock {\em Phys. Rev.} A {\bf 68} 022111

\bibitem{bn2003b}
Navarro B, Egusquiza I~L, Muga J~G and Hegerfeldt G~C 2003
\newblock {\em J. Phys.} B {\bf 36} 3899

\bibitem{dambo2003}
Damborenea J~A, Egusquiza I~L, Hegerfeldt G~C and Muga J~G 2003
\newblock {\em J. Phys. B: At. Mol. Opt. Phys.} {\bf 36} 2657

\bibitem{hsmn2004}
Hegerfeldt G~C, Seidel D, Muga J~G and Navarro B 2004
\newblock {\em Phys. Rev.} A {\bf 70} 012110

\bibitem{rdnmh2004}
Ruschhaupt A, Damborenea J~A, Navarro B, Muga J~G and Hegerfeldt G~C 2004
\newblock {\em Europhys. Lett.} {\bf 67} 1

\bibitem{hhm2005}
Hannstein V, Hegerfeldt G~C and Muga J~G 2005
\newblock {\em J. Phys.} B {\bf 38} 409

\bibitem{bf2002}
Brunetti R and Fredenhagen K 2002
\newblock {\em Phys. Rev.} A {\bf 66} 044101

\bibitem{qujump}
Hegerfeldt G~C and Wilser T~S 1992 \emph{Classical and Quantum
  Systems. Proc. of the 2nd Int. Wigner Symp., July
  1991} ed H~D Doebner, W Scherer and F Schroeck (Singapore: World
Scientific) p~104; Hegerfeldt G~C 1993 {\em Phys. Rev.} A {\bf 47}
449; Hegerfeldt G~C and Sondermann D 1996 {\em Quant. Semiclass. Opt.}
{\bf 8} 121. For a review cf.\ \cite{pk1998}. The quantum jump
approach is essentially equivalent to the Monte-Carlo wavefunction
approach \cite{dcm1992}, and to the quantum trajectories of H
Carmichael \cite{carm1993}.

\bibitem{beskow}
Beskow A and Nilsson J 1967
\newblock {\em Arkiv f\"or Fysik} {\bf 34} 561

\bibitem{khalfin}
Khalfin L~A 1968
\newblock {\em Zh.\ Eksp.\ Teor.\ Fiz.\ Pis.\ Red.} {\bf 8} 106
\newblock [JETP Lett.\ {\bf 8} 65 (1968)]

\bibitem{Sudarshan}
Misra B and Sudarshan E~C~G 1977 J. Math. Phys. \textbf{18} 756. For
  further references see \cite{bh1996}.

\bibitem{vN}
von Neumann J 1932
\newblock {\em Mathematische Grundlagen der Quantenmechanik}.
\newblock (Berlin: Springer)
\newblock {[}english translation: von Neumann J 1955
\emph{Mathematical Foundations of Quantum Mechanics} (Princeton:
Princeton University Press){]} chapter V.1

\bibitem{Lueders}
L{\"u}ders G 1951
\newblock {\em Ann. Phys. (Leipzig)} {\bf 443} 322

\bibitem{gch2003}
Hegerfeldt G~C 2003
\newblock {\em Lecture Notes in Physics} vol 622 {\em Irreversible
  Quantum Dynamics} ed F Benatti and R Floreanini
(Berlin Heidelberg: Springer) p~233

\bibitem{mbm1995}
Muga J~G, Brouard S and Mac{\'i}as D 1995
\newblock {\em Ann. Phys. (NY)} {\bf 240} 351

\bibitem{pms1998}
Palao J~P, Muga J~G and Sala R 1998
\newblock {\em Phys. Rev. Lett.} {\bf 80} 5469

\bibitem{nhs}
Hegerfeldt G~C, Neumann J~T and Schulman L~S 2006
\newblock Passage-time distributions from a spin-boson detector model
\newblock {\em Preprint} quant-ph/0610041

\bibitem{ap1980}
Peres A 1980
\newblock {\em Am. J. Phys.} {\bf 48} 552

\bibitem{amm2003}
Alonso D, {Sala Mayato} R and Muga J~G 2003
\newblock {\em Phys. Rev.} A {\bf 67} 032105

\bibitem{pk1998}
Plenio M~B and Knight P~L 1998
\newblock {\em Rev. Mod. Phys.} {\bf 70} 101

\bibitem{dcm1992}
Dalibard J, Castin Y and M\o{}lmer K 1992
\newblock {\em Phys. Rev. Lett.} {\bf 68} 580

\bibitem{carm1993}
Carmichael H 1993 \newblock
{\em Lecture Notes in Physics} vol m 18 {\em An Open System Approach
  to Quantum Optics} 
\newblock (Berlin Heidelberg: Springer Verlag)

\bibitem{bh1996}
Beige A and Hegerfeldt G~C 1996
\newblock {\em Phys. Rev.} A {\bf 53} 53

\end{thebibliography}
\end{document}